\documentclass[%
preprint,
 amsmath,amssymb,
 aps,
]{revtex4-1}

\usepackage{amsmath}
\usepackage{graphicx}
\usepackage{subfig}
\usepackage{dcolumn}
\usepackage{bm}
\usepackage{extarrows}
\usepackage{multirow}

\begin{document}

\title{Predicting acoustic relaxation absorption in gas mixtures for extraction of composition relaxation contributions}

\author{
Tingting Liu, Shu Wang and Ming Zhu}
\email{zhuming@mail.hust.edu.cn}
\address{School of Electronic Information and Communications, Huazhong University of Science and Technology,
Wuhan 430074, P.R. China\\
}











\begin{abstract}
The existing molecular relaxation models based on both parallel relaxation theory and series relaxation theory cannot extract the contributions of gas compositions to acoustic relaxation absorption in mixtures. In this paper, we propose an analytical model to predict acoustic relaxation absorption and clarify composition relaxation contributions based on the rate-determining energy transfer processes in molecular relaxation in excitable gases. By combining parallel and series relaxation theory, the proposed model suggests that the vibration-translation process of the lowest vibrational mode in each composition provides the primary deexcitation path of the relaxation energy, and the rate-determining vibration-vibration processes between the lowest mode and others dominate the coupling energy transfer between different modes. Thus, each gas composition contributes directly one single relaxation process to the molecular relaxation in mixture, which can be illustrated by the decomposed acoustic relaxation absorption spectrum of the single relaxation process. The proposed model is validated by simulation results in good agreement with experimental data such as  $\mathrm{N_2}$,  $\mathrm{O_2}$,  $\mathrm{CO_2}$,  $\mathrm{CH_4}$ and their mixtures.
\end{abstract}


\maketitle
\section{Introduction}
In gases with internal degrees of freedom, acoustic waves show anomalous absorption in excess of the classical absorption over the range of low and moderate sound frequency. This phenomenon is ascribed to the molecular relaxation involved with the adjustment rate of energy exchange between the internal and external degrees of freedom.\cite{herzfeld1928dispersion,borrell1967relaxation,ben1974vibrational,bhattacharjee1981dynamic,hancock2009} Thus the molecular energy transfer mechanism lays the significant foundation for the molecular relaxation, and determines the acoustic relaxation absorption. Developed with different molecular energy transfer mechanisms, existing relaxation theories fall into two categories, parallel relaxation theory and series relaxation theory.\cite{herzfeld1959absorption} However, both of them cannot extract the contributions of gas compositions to relaxation absorption of sound waves.



In the parallel relaxation theory for sound propagation in gases, each vibrational mode of molecules provides a deexcitation path with a characteristic relaxation time for the resident energy in internal degrees of freedom.\cite{lambert1977vibrational} The parallel relaxation theory has been studied extensively to predict the acoustic relaxation absorption in excitable gases (including diatomic and polyatomic gases and mixtures). Schwartz, Slawsky and Herzfeld (SSH) \cite{schwartz1952calculation} described the single relaxation processes in diatomic gases and then the SSH theory was extended to predict the multiple relaxation processes in polyatomic gases and gas mixtures.\cite{tanczos1956calculation,bauer1972multimode,Dain2001,petculescu2005fine,ke2013decoupling,liu2017decomposition} Based on the SSH parallel relaxation equations, Bauer\cite{bauer1965phenomenological} and Zhang\cite{zhang2017calculating} decoupled a multi-relaxation process into the sum of single-relaxation processes. The decoupled results show that only the lowest vibrational modes of gas compositions provide predominant contributions to acoustic absorption. However, each of the decoupled single processes contains the contributions from all the vibrational modes. Therefore, the parallel relaxation theory needs to be improved to clarify the relaxation contributions of gas compositions.

In the series relaxation theory, another phenomenological theory, only the lowest vibrational mode of a molecule is directly excited by the sound wave and the energy flows rapidly from this mode into others.\cite{lambert1977vibrational} The lowest mode provides the only deexcitation path for the relaxation energy in molecule, and the coupling energy transfer processes between different modes are not considered due to the rapid energy flow. Thus the relaxation process of pure gas is characterized by a single relaxation process with single relaxation time. The acoustic relaxation absorption predicted by series theory is verified for pure gases by the experimental results from Shields,\cite{shields1957thermal,shields1971more,shields1980vibrational}, Henderson,\cite{henderson1959ultrasonic} and Gravitt.\cite{gravitt1966thermal} Nevertheless, the series relaxation theory ignores the interaction effects between vibrational modes of different components, which makes them only applicable for pure gases.

In this paper, we develop an analytical model of molecular relaxation to predict acoustic relaxation absorption and clarify the contributions of gas compositions by combining the parallel and series relaxation theory. In Section 2, we obtain the analytical model for acoustic relaxation absorption based on the rate-determining relaxation energy transfer processes. In Section 3, the proposed model is validated with experimental data. In our model, each gas composition contributes directly one single relaxation process to molecular relaxation of mixture. For example, the two peaks on relaxation absorption spectrum of $\mathrm{CO_2-O_2}$ are contributed from single relaxation processes of $\mathrm{CO_2}$ and $\mathrm{O_2}$, respectively. Conclusions are given in Section 4.

\section{ Analytical model for acoustic relaxation absorption}
\subsection{Rate-determining energy transfer processes in molecular relaxation of gases}

Upon the passage of sound wave in excitable gases, acoustic wave disrupts the gas molecular equilibrium, and a redistribution of molecular energy among internal (vibrational and rotational) and external (translational) degrees of freedom occurs to reach a new equilibrium.\cite{parker1959rotational,zhang2008approach,panesi2014nonequilibrium} For most common gaseous molecules at ordinary temperature, molecular rotation adjusts instantaneously to energy changes of translational degree of freedom, and molecular vibration with large quantum level spacing is unable to follow the translational temperature fluctuations.\cite{fernandes2004sound,li2016rotational} As a result, the molecular vibrational energy transfer referred as the thermal molecular relaxation gives rise to relaxation absorption of acoustic energy.

\subsubsection{Rate-determining V-V and V-T transfer processes}

There are two kinds of vibrational energy exchange processes during bimolecular collisions, vibration-translation (V-T) and vibration-vibration (V-V) transfer processes.\cite{flynn1996vibrational,aliat2011simple,wu2012kinetic} The V-T and V-V processes can be expressed by the transfer equations as follows: \cite{Dain2001}
\begin{align}
 A^*(p+1)+X(q)&\xLongleftrightarrow {V-T} A(p)+X(q)+\Delta E,
\tag{a}\\
 A^*(p+1)+X(q)&\xLongleftrightarrow {V-V} A(p)+X^*(q+1)+\Delta E.
\tag{b}
\end{align}
where $A$ and $X$ represent collision gas molecules of the same or different species, $p$ and $q$ represent molecular vibrational modes, the asterisk indicates a molecule excited to the higher vibrational level, and $\Delta E$ is the exchanged energy during the collision process. The V-T process results in energy transferring from molecular vibrational modes to molecular translation, i.e. one molecule loses one quantum while the vibrational state of the other molecule is unaltered. The V-V process leads to the energy transition of different vibrational modes, and involves with simultaneous changes of vibrational states in both of the collision molecules.

The molecular energy transfer processes in which the molecules change states upon collisions can be described analytically based on quantum mechanical theory.\cite{gao1998quantum,mattioli2015classical,gutierrez2016non,mammoli2016collisional} During the bimolecular collision of gas species $a$ with species $b$, transition probability $P^{i_a-j_a}_{i_b-j_b}(a,b)$ that a pair of molecules originally with vibrational states $i_a$ and $i_b$ will arrive at vibrational states $j_a$ and $j_b$ after collision, is derived from the Tanczos equation,\cite{tanczos1956calculation}
\begin{multline} \label{eq1}
P^{i_a-j_a}_{i_b-j_b}(a,b)=P_0(a)P_0(b)(\frac{\gamma_c^*}{\sigma})^2[V(a)]^2[V^(b)]^2 8(\frac{\pi}{3})^{1/2}\\
[\frac{8\pi^3\mu \Delta E}{(\alpha^*h)^2}]^2 \zeta^{1/2}\exp(-3\zeta+\frac{\Delta E}{2k_B T}+\frac{\varepsilon}{k_B T}),
\end{multline}
where $\Delta E=h\nu_a(i_a-j_a)+h\nu_b(i_b-j_b)$ is the exchanged energy during the collision processes, $h$ is Planck constant and $\nu_a$, $\nu_b$ are the vibrational frequencies of species $a$ and species $b$. The molecule usually gains or loses one quantum because the molecular population of the higher excited states is so small at ordinary temperatures, i.e. $i$ and $j=$0 or 1.\cite{dmitrieva1990vt} Since $\Delta E$ is the amount of vibrational energy transferred in the collision, $\Delta E$ is usually expressed by the difference values of vibrational frequencies without the constant $h$.\cite{lisak2009spectroscopic} The detail parameters in equation (2.1) are not introduced here and some parameters values for common gases are listed in Table 1.\cite{lambert1977vibrational,ke2012analytical,petculescu2005fine}

The number of collisions made by one molecule per second $Z(a,b)$ can be written as\cite{lambert1977vibrational},
\begin{equation} \label{eq2}
Z(a,b)=2 N_b(\frac{\sigma_a+\sigma_b}{2})^2(\frac{2\pi kT (m_a+m_b)}{m_a m_b})^{1/2},
\end{equation}
where $N_b$ is the number of molecules per unit volume, $\sigma_a$ and $\sigma_b$ are the collision diameters, $m_a$ and $m_b$ are the molecular mass of species. Combing equations (2.1) and (2.2), the rate of energy transfer process $\beta$, which is inversely proportional to the number of effective collisions that a molecule experiences per second, is obtained as:\cite{schwartz1954vibrational}
\begin{equation} \label{eq3}
\beta^{i_a-j_a}_{i_b-j_b}=\frac{1}{Z(a,b)P^{i_a-j_a}_{i_b-j_b}(a,b)},
\end{equation}
In equation (2.3), the transfer rate $\beta$ connects the microscopic gas-kinetic properties with the macroscopic molecular energy transfer mechanism. The transfer rates of V-T and V-V processes greatly influence the vibrational energy transfer of molecular relaxation process, especially, the relaxation time.\cite{nekouzadeh2005wave,hancock2006vibrational,few2014rate} From equations (2.1), (2.2) and (2.3), the smaller exchanged energy needs less kinetic energy for energy transfer in collisions, resulting in larger transition probabilities and faster energy transfer of molecular relaxation process, i.e. faster relaxation times.\cite{suraud2014non,dashevskaya2015vibrational} Therefore, we investigate the influence of transfer rates of V-T and V-V processes on molecular relaxation and suggest the concept of rate-determining energy transfer process based on the exchanged energy $\Delta E$.

\begin{table*}[htbp]
\caption{Parameter values of gases for the calculation of transition probabilities}
\label{Tab1}
\centering
\begin{tabular}{p{30pt}p{60pt}p{50pt}p{60pt}p{60pt}p{60pt}p{55pt}p{40pt}}
\hline
Gas & Molecular weight $m$ & normal modes of vibration $\nu$ (cm$^{-1}$)
& Collision diameter $\delta$ (10$^{-10}$m)   & Depth of potential well $\epsilon/k_B$(K)  &Vibrational amplitude coefficient $V^2$ (amu$^{-1}$) & Degeneracy $g$ & Steric factor $P_0$ \\
\hline
\multirow{3}{*}{$\mathrm{CO_2}$}  & \multirow{3}{*}{44}  & 667 &  \multirow{3}{*}{3.996} &\multirow{3}{*}{190.32}  & \multirow{3}{*}{0.05} & 2 &\multirow{3}{*}{2/3} \\
 & & 1388 & & & &1 & \\
 & & 2349 & & & &1 & \\
 \hline
\multirow{4}{*}{$\mathrm{CH_4}$}  & \multirow{4}{*}{16}  & 1306 &  \multirow{4}{*}{4.075} &\multirow{4}{*}{143.91}  & 0.8368 & 3 &\multirow{4}{*}{2/3} \\
 & & 1534 & & &0.9921 &2& \\
 & & 2915 & & &0.9921 &1& \\
 & & 3019 & & &0.9923 &1& \\
 \hline
$\mathrm{N_2}$  & 28 & 2331  &3.074 &80.01  & 0.0714 & 1 &1/3\\
\hline
$\mathrm{O_2}$  & 32 & 1554  &3.541 &88.17  & 0.0625 & 1 &1/3\\
\hline
\end{tabular}
\end{table*}

In figure 1, a simple example of gas molecule with two vibrational modes is adopted here to illustrate the influence of the energy transfer rate. Figure 1 depicts the V-T energy transfer between the lowest vibrational mode $\nu_1$ and the ground state $\nu_0$, and the V-V transfer between $\nu_1$ and higher mode $\nu_2$. The V-T process of the vibrational mode provides the deexcitation path for energy transfer of the mode and the V-V process describes energy transfer between different vibrational modes. In our work, special attentions are paid to the rate-determining V-V process to describe the V-V processes which determine the vibrational energy transfer between different modes in molecular relaxation process. Since the exchanged energy $\Delta E$ between the two vibrational modes will greatly influence vibrational energy transfer rate of the V-V process,\cite{shields1973acoustical} the rate-determining V-V process is defined by the exchanged energy $\Delta E$. As illustrated in figure 1, the exchanged energy of the rate-determining V-V process ranges from $\Delta E_{min}$ to $\Delta E_{max}$. For example, the energy gap between $\nu_1$ and $\nu_2$ in figure 1 is within the range from $\Delta E_{min}$ to $\Delta E_{max}$, thus the V-V energy transfer process between them would play significant role in the molecular energy relaxation and the process is regarded as rate-determining V-V process.

\begin{figure} [htbp]
\label{figure1}
\centering
\includegraphics
[width=0.8\textwidth,height=0.65\textwidth]{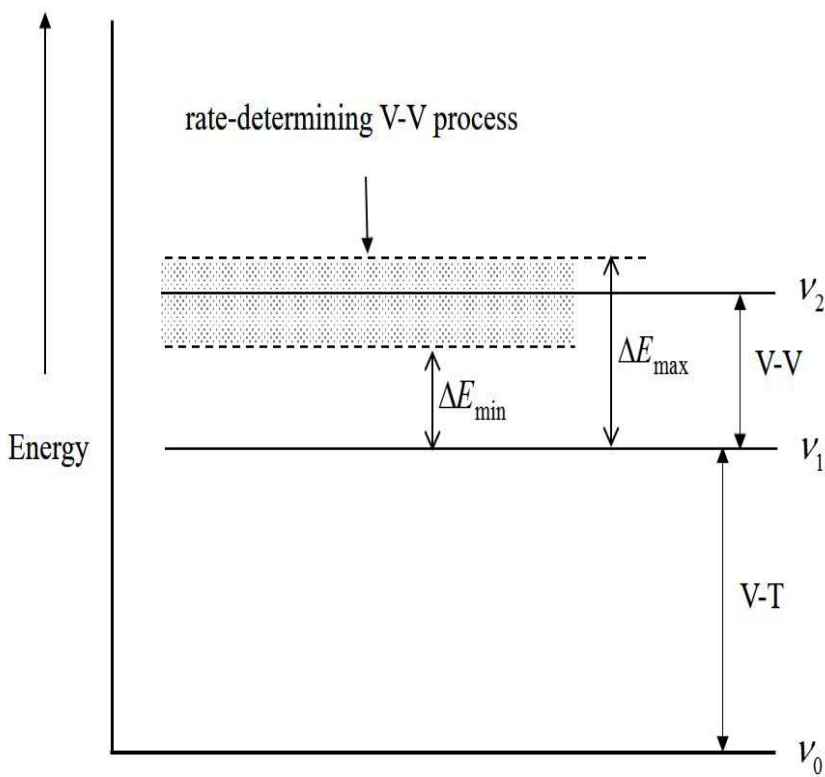}
\caption{\label{fig:epsart} Rate-determining V-V and V-T energy transfer processes due to different exchanged energy $\Delta E$ for a simple molecule with two vibrational modes}
\end{figure}

For a certain vibrational mode, the rate-determining V-V process within the range from $\Delta E_{min}$ to $\Delta E_{max}$ dominates in energy transfer between the mode and others. Provided that the exchanged energy is smaller than $\Delta E_{min}$ or greater than $\Delta E_{max}$, the V-V processes will not dominate the energy transfer. In the first case, the energy gap between two vibrational modes, which is smaller than $\Delta E_{min}$, causes very rapid vibrational energy transfer. Then the rapid V-V energy transfer process would keep energy equilibrium between the two vibrational modes, such as resonant transition.\cite{rapp1964resonant} So the rapid V-V process will not be considered for energy transition of molecular relaxation. In the second case, the large energy gap results in slow energy transfer which has little influence on molecular relaxation. And this case occurs for the V-V processes with exchanged energy greater than the value $\Delta E_{max}$. Therefore, for a vibrational mode, the V-V processes within the range from $\Delta E_{min}$ to $\Delta E_{max}$ are the rate-determining processes. According to the previous investigations\cite{brenner1981near,shin1984role,shin1986vibrational,bohn1999vibrational} and our simulations, we set the $\Delta E_{min}$ and $\Delta E_{max}$ as values 100 cm$^{-1}$ and 200 cm$^{-1}$, by the difference values of vibrational frequencies. Next, we consider the vibrational energy exchange in gases with multiple vibrational modes and focus on the transfer process of the lowest vibrational mode.

\subsubsection{Rate-determining transfer processes of gases with multiple vibrational modes}

For polyatomic gases and gas mixtures composed of multiple vibrational modes, there arises varieties of V-V and V-T processes. The traditional parallel relaxation theory considers the energy transfer between all the vibrational modes. According to the decomposition in Ref. \cite{zhang2017calculating}, the vibrational modes except the lowest one of a gas composition, provide negligible contributions to the acoustic relaxation absorption in terms of the decoupling results for most common gases, such as  $\mathrm{N_2}$, $\mathrm{O_2}$, $\mathrm{CO_2}$, $\mathrm{CH_4}$ and so on. Thus we focus on the transfer process of the lowest vibrational mode in this work. The traditional series relaxation theory suggests almost all the vibrational energy relaxes via the lowest mode, and only the V-T process of the lowest vibrational mode is taken into account for molecular relaxation, which is not accurate. According to our discussion on energy transfer processes, the energy transfer of the lowest vibrational mode includes the V-T process and V-V process. The V-T process of the lowest vibrational mode provides the primary deexcitation path for vibrational energy transfer of molecule, and the rate-determining V-V processes are regarded as the main paths of molecular energy transfer between the lowest one and the others. Therefore, the rate-determining V-V process and the V-T process of the lowest vibrational mode would play significant roles in molecular energy relaxation and influence the macroscopic relaxation characteristics.

Based on the proposed energy transfer mechanism, we provide an analytical model of the molecular relaxations for pure gases and gas mixtures. In our model, energy transfer of pure gases is realized by the rate-determining V-V process and the V-T process of the lowest vibrational mode, and the molecular relaxation of multi-component gas mixtures is achieved by the energy transfer process of each component. Based on the analytical model, the acoustic relaxation absorption of pure gases and mixtures could be calculated.

\subsection{Prediction of acoustic relaxation absorption in pure gases and mixtures}

In the studies of acoustic relaxation, the complex-valued and frequency-dependent effective specific heat is the macroscopic footprint of molecular thermal relaxation\cite{petculescu2005synthesizing} and the essential parameter to calculate the acoustic relaxation absorption. In this section, the expressions of effective specific heats are obtained and acoustic relaxation absorption is calculated for molecular relaxation of pure gases and mixtures based on the proposed energy transfer mechanism.

\subsubsection{Acoustic relaxation of pure gases}

For the simplest case of pure diatomic gases, only the V-T process occurs and the V-V process does not appear due to the one vibrational mode in the diatomic molecule. The only vibrational mode is directly excited by the sound wave and relaxation energy flows from this mode into the translation via the V-T process. The V-T deexcitation of the vibrational energy could be described by the equation of complex-valued and frequency-dependent effective specific heat,
\begin{equation} \label{eq4}
C_V^{eff}(\omega)=C_V^\infty+\frac{C_V^{vib}}{1+i\omega\tau}
\end{equation}
where $C_V^\infty$ is the external specific heat, $C_V^{vib}$ is the vibrational specific heat calculated from the Planck-Einstein function, $\tau$ represents the relaxation time,\cite{bauer1965phenomenological}
\begin{equation} \label{eq5}
\tau^{-1}=(\tau^{tran})^{-1}=\beta^{1-0}_{0-0}(1-\exp (h\nu /{k_B T_0})),
\end{equation}
where $\tau^{tran}$ represents the deexcitation time of V-T process and is called translational relaxation time, $\beta^{1-0}_{0-0}$ represents the rate of V-T process from equation (2.3), $\nu$ is vibrational frequency of the molecule, $k_B$ is Boltzmann constant and $T_0$ is equilibrium temperature.

For a pure gas composed of multiple vibrational modes, the V-T process and the rate-determining V-V process of the lowest mode play significant roles in molecular energy relaxation. The V-T process provides the path for energy transferring from the lowest mode to the translation, while the rate-determining V-V processes are the dominant paths of energy transfer between the lowest one and the others. Since polyatomic pure gases show a single vibrational relaxation process,\cite{herzfeld1959absorption} the frequency-dependent effective specific heat of this relaxation process is characterized by one single relaxation time as,
\begin{equation} \label{eq8}
C_V^{eff}(\omega)=C_V^\infty+\frac{C^{int}}{1+i\omega\tau^{eff}}
\end{equation}
where $C^{int}$ is the whole internal specific heat, $C^{int}=\sum_{j=1}^N C_j^{vib} $, $C_j^{vib}$ is the vibrational specific heat of vibrational mode $j$ and $N$ represents the number of molecular vibrational modes, $\tau^{eff}$ is the effective relaxation time as\cite{herzfeld1928dispersion}
\begin{equation} \label{eq9}
\tau^{eff}=\tau_1 \frac{C^{int}}{C_1^{vib}}.
\end{equation}
Here $\tau_1$ represents the relaxation time of the lowest mode of the gas molecule, and $C_1^{vib}$ is the vibrational specific heat of lowest vibrational mode.

The calculation of the time $\tau_1$ is based on the kinetic theory for energy transfer process. The energy transfer of the lowest vibrational mode includes the V-T process where a single quantum is exchanged with translation, and V-V process where collision molecular partners gain or lose one quantum as shown in (a) and (b). Thus the relaxation time of the lowest vibrational modes can be expressed with the times for V-T and V-V relaxation processes as,
\begin{equation} \label{eq10}
\frac{1}{\tau_1}=\frac{1}{\tau_1^{tran}}+\frac{1}{\tau_1^{vib}}.
\end{equation}
The translational relaxation time $\tau_1^{tran}$ is calculated based on the deexcitation time of V-T process. When calculating vibrational relaxation time $\tau_1^{vib}$, only the rate-determining V-V processes play significant role in molecular energy transfer. Since the rate-determining V-V processes are defined by the exchanged energy, we take into account of the relaxation time of rate-determining V-V processes with exchanged energy within a range of 100 cm$^{-1}< |\Delta E| <$200 cm$^{-1}$. Thus the translational relaxation time $\tau_1^{tran}$ and the vibrational relaxation time $\tau_1^{vib}$ are expressed as follows.
\begin{align} \label{eq11}
(\tau_1^{tran})^{-1}&= \beta_{0-0}^{1-0}(1,1)(1-\exp (h\nu_1/{k_B T_0})),
\\
(\tau_1^{vib})^{-1}&=\sum_{j=2}^{N} \beta_{0-1}^{1-0}(1,j),j\neq 1, j\mbox{ decided by } \Delta E.
\end{align}
Substituting equations (2.9) and (2.10) into equation (2.8), the relaxation time of the lowest vibrational mode could be obtained. Hence, the effective specific heat in equation (2.6), characterized by a single relaxation time, is expressed for molecular thermal relaxation for pure polyatomic gases.

\subsubsection{Acoustic relaxation of gas mixtures}

Based on the fact that the energy transfer process of each component constitutes the molecular relaxation of a gas mixture, the acoustic relaxation of mixture is calculated from the superposition of each gas composition contribution. For the gas mixtures with M gas components, its effective specific heat is calculated by summing all the specific heat contributions of gas compositions,
\begin{equation} \label{eq12}
C_V^{eff}(\omega)=C_m^\infty+\sum_{k=1}^M {\frac{a_k C_k^{int}}{1+i\omega \tau_k^{eff}}},
\end{equation}
where $C_m^\infty$ is the external specific heat of mixture, $C_m^\infty=\sum_{k=1}^{M} a_k C_k^{\infty}$; $a_k$ is the mole fraction of gas molecule $k$; $C_k^{int}$ is the internal specific heat of gas molecule $k$ ( $C_k^{int}=\sum_{j=1}^{N_k} C_j^{vib}$) and $\tau_k^{eff}$ is the effective relaxation time of gas molecule $k$. Similarly with equation (2.7) for pure gases, $\tau_k^{eff}$ is determined by the relaxation time of the lowest vibrational mode of gas molecule $k$,
\begin{equation} \label{eq13}
\tau_k^{eff}=\tau_{k_1} \frac{C_k^{int}}{C_{k_1}^{vib}},
\end{equation}
where the subscript $k_1$ denotes the lowest vibrational mode of gas molecule $k$, $\tau_{k_1}$ is the relaxation time of the lowest vibrational mode in molecule $k$, and $C_{k_1}^{vib}$ is the vibrational specific heat of the lowest mode. Similarly with the calculation for polyatomic molecules in equation (2.8), $\tau_{k_1}$ is calculated from the relaxation times of V-T and V-V processes of the lowest mode.
\begin{equation} \label{eq14}
\frac{1}{\tau_{k_1}}=\frac{1}{\tau_{k_1}^{tran}}+\frac{1}{\tau_{k_1}^{vib}},
\end{equation}
where
\begin{align}\label{eq15}
(\tau_{k_1}^{tran})^{-1}&=\sum_{l=1}^M a_k \beta_{0-0}^{1-0}(k_1,l)(1-\exp (h\nu_{k_1}/{k_B T_0})),\\
(\tau_{k_1}^{vib})^{-1}&=\sum_{j=1}^{N} a_k \beta_{0-1}^{1-0}(k_1,j), j\neq k_1, j \mbox{ decided by} \Delta E.
\end{align}
The translational relaxation time $\tau_{k_1}^{tran}$ is the V-T deexcitation time of the lowest mode in gas molecule $k$. In the V-T transitions, the lowest mode exchanges single quantum with translation and the energy of molecule $l$ keeps unchanged.\cite{ke2013decoupling} The vibrational relaxation time $\tau_{k_1}^{vib}$ takes into account of the V-V energy transfer processes between the lowest mode and other modes $j$ with exchanged energy within a range of 100 cm$^{-1}< |\Delta E| <$200 cm$^{-1}$.

Using equation (2.12), the effective relaxation times for each gas component of mixture are obtained respectively. The frequency-dependent effective specific heat $C_V^{eff}$ is calculated for gas mixture by the superposition of containing gas composition contributions in equation (2.11). Consequently, we obtain the effective specific heat of molecular relaxation of gas mixtures to calculate acoustic relaxation absorption.

\subsubsection{Acoustic relaxation absorption of gases}

During the propagation of acoustic waves in excitable gases, measuring the frequency-dependent acoustic relaxation absorption is an effective method to characterize molecular relaxation process. Hence, we calculate the acoustic relaxation absorption per wavelength $\alpha_r\lambda$ to observe the molecular relaxation characteristics from a macroscopic perspective.
The sound propagation in gases can be described by the complex-valued effective acoustic wave number $\tilde{k}$ which relates the thermodynamic effective specific heat and the macroscopic gas properties together as,
\begin{equation} \label{eq16}
\tilde{k}(\omega)=\frac{\omega}{c(\omega)}-i\alpha_r(\omega)=\omega \sqrt{\frac{\rho_0}{p_0}}\sqrt{\frac{C_V^{eff}(\omega)}{C_V^{eff}(\omega)+R}},
\end{equation}
where $\rho_0$ and $p_0$ are the equilibrium gas density and pressure, respectively, R is the universal gas constant. The real and imaginary parts of $\tilde{k}(\omega)$ yield the frequency-dependent sound speed $c$ and the relaxational absorption $\alpha_r$, respectively. With the relations in equation (2.16), we can obtain the acoustic relaxation absorption per wavelength as
\begin{equation} \label{eq17}
\mu(\omega)=\alpha_r(\omega) \lambda= \alpha_r(\omega) c(\omega)/f
\end{equation}
where $\lambda$ represents sound wavelength, $\mu$ is called acoustic relaxation absorption spectrum. As a result, the relaxation absorption of acoustic wave in pure gases and mixtures can be predicted based on the expressions of effective specific heat from our model.

Consequently, we propose an analytical model to predict the molecular relaxation absorption in gases. Compared with traditional parallel relaxation theory, the mechanism of energy transfer between all the vibrational modes is clarified in this work, due to the main role of molecular lowest vibrational mode in energy relaxation. Compared with traditional series relaxation theory, our work considers the influence of rate-determining V-V process on the energy transfer and develops a complete analytical expression for molecular relaxation absorption in gases.

\section{Results and discussion}
\subsection{Results of acoustic relaxation absorption spectra}

In molecular acoustic experiments, the absorption of sound wave is measured over a wide frequency range, which is illustrated by the acoustic absorption spectra. The total sound absorption is the sum of the relaxational part and the classical part from transport phenomena. The classical absorption generally plays a major role at the frequencies in megahertz and higher, while the relaxation absorption is dominant at low and moderate frequencies. Here we obtain theoretical acoustic absorption spectra including the relaxation part calculated from our model and the classical absorption calculated using the formulation of Stokes and Kirchhoff.\cite{bass2001absorption} The theoretical spectra are compared with the experimental results to validate our proposed model.

\begin{figure} [htbp]
\label{figure2}
\includegraphics
[width=0.8\textwidth,height=0.65\textwidth]{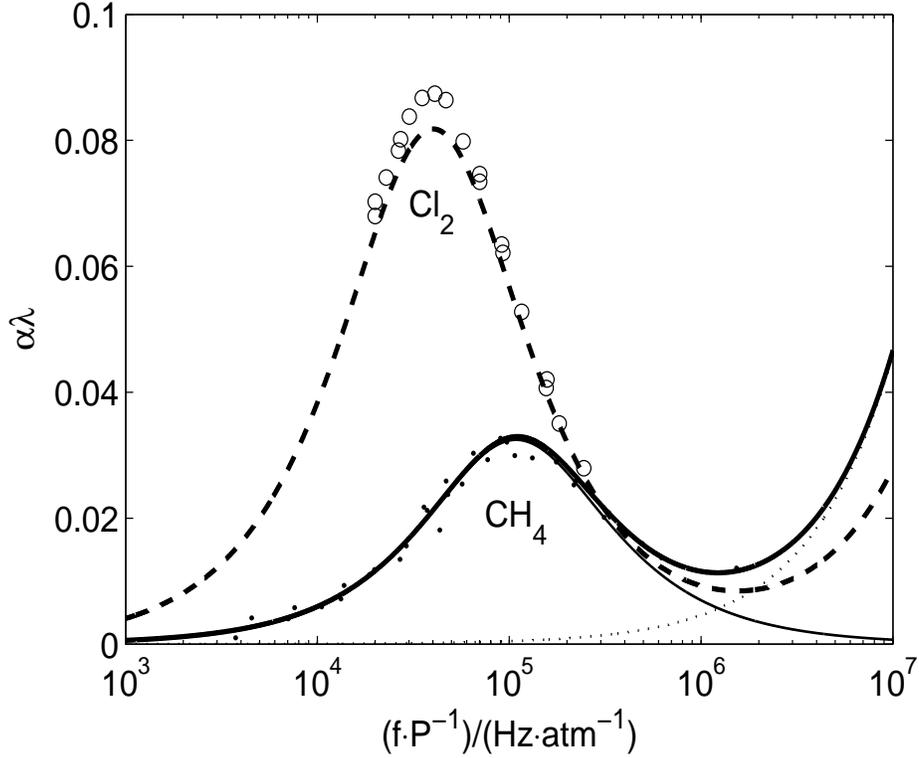}
\caption{\label{fig:epsart} Comparisons of theoretical acoustic absorption spectra predicted from our model with experimental data for pure gases of $\mathrm{Cl_2}$ and $\mathrm{CH_4}$. Curves: Theoretically predicted total absorption (relaxation and classical absorption) spectra for $\mathrm{CH_4}$ at T= 293.9 K (solid curves) and for $\mathrm{Cl_2}$ at T= 296 K (dashed curves); theoretical predicted relaxation absorption (fine curve) and classical absorption (dotted curve) for $\mathrm{CH_4}$ at T= 293.9 K. Symbols: Experimental data for $\mathrm{Cl_2}$ from Shields\cite{shields1960sound} (circles) and for $\mathrm{CH_4}$ from Ejakov\cite{ejakov2003acoustic} (dots).}
\end{figure}

\begin{figure} [htbp]
\label{figure3}
\includegraphics
[width=0.8\textwidth,height=0.65\textwidth]{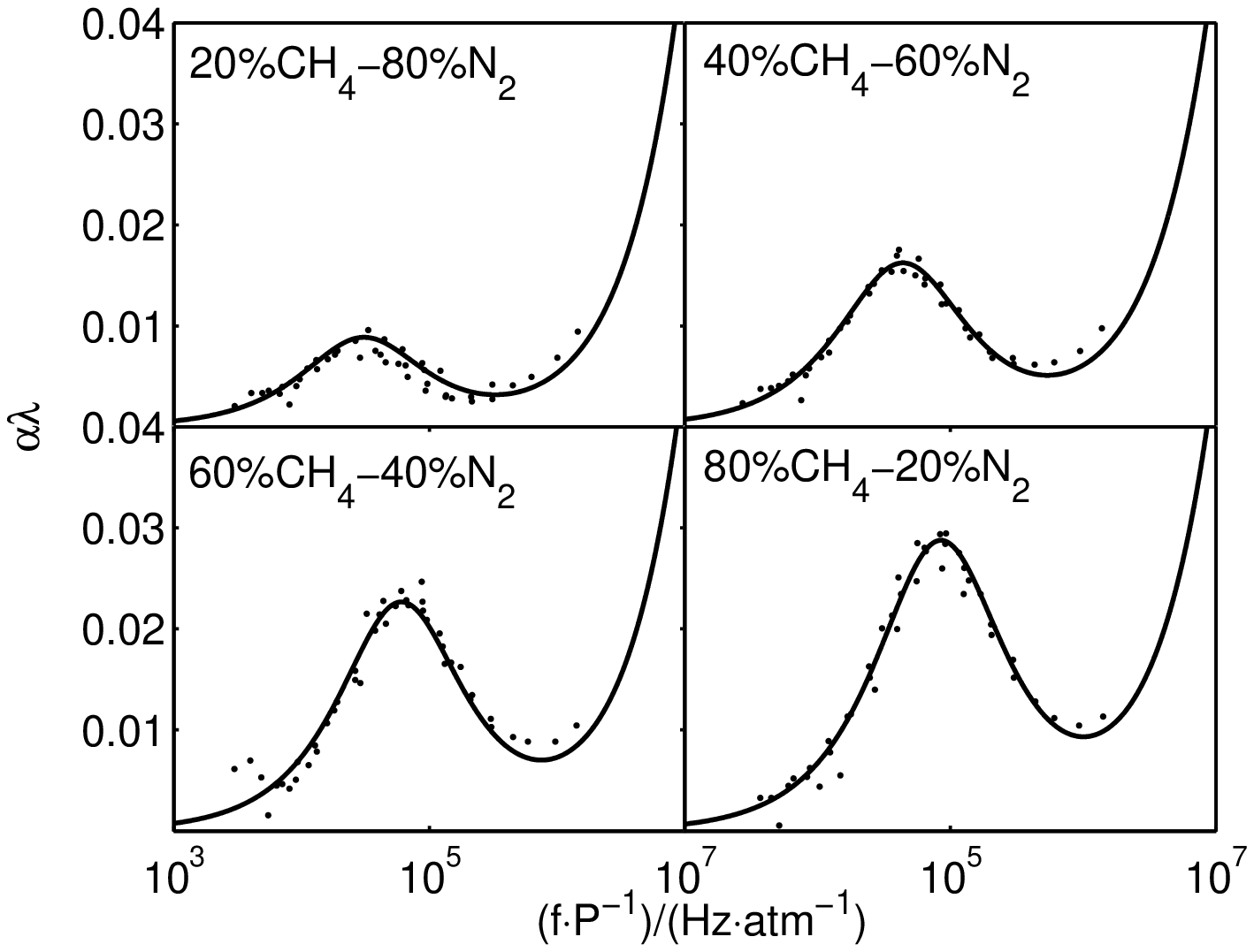}
\caption{\label{fig:epsart} Comparisons of theoretical acoustic absorption (relaxation and classical absorption) spectra  predicted from our model (solid curves) with experimental data from Ejakov\cite{ejakov2003acoustic}(dots) in $\mathrm{CH_4-N_2}$ mixtures with 20$\%$, 40$\%$, 60$\%$, 80$\%$ $\mathrm{CH_4}$ at temperatures of 293.2, 293.0, 293.4 and 295 K, respectively.}
\end{figure}

In figure 2 and 3, the comparisons of absorption spectra for pure gases and gas mixtures are provided. In figure 2, the theoretical total absorption spectra of pure $\mathrm{Cl_2}$ (diatomic gas) and pure $\mathrm{CH_4}$ (polyatomic gas) match well with the experimental data from Shields\cite{shields1960sound} and Ejakov\cite{ejakov2003acoustic}. For gas mixtures $\mathrm{CH_4-N_2}$ with different concentrations in figure 3, the spectral curves predicted from our model also agree well with the experimental results from Ejakov's work\cite{ejakov2003acoustic}. Due to the classical absorption dominant at $f/p>$1 MHz/atm, the absorption curves show ascending tendency at high frequency. The results validate the our proposed model in the prediction of acoustic relaxation absorption.

\begin{table}[htbp]
\caption{Comparisons of theoretical calculations from our proposed model and experimental data for the relaxation frequencies of various gases at room temperature}
\label{Tab2}
\centering
\begin{tabular}{p{95pt}p{60pt}p{65pt}p{55pt}p{55pt}}
\hline
Gas & Theoretical relaxation frequency (Hz) & Experimental relaxation frequency (Hz)
& Tempera- ture (K)   & Reference \\
\hline
$\mathrm{O_2}$  & 1.08$\times10^1$  &$\sim$1$\times10^1$     &303.2  &\cite{holmes1963vibrational}\\
$\mathrm{CH_4}$ & 1.46$\times10^5$ &$\sim1.6\times10^5$  &299.15 &\cite{gravitt1966thermal}\\
$\mathrm{CO_2}$ & 4.11$\times10^4$ &$\sim$3.9$\times10^4$  &296.15 &\cite{angona1953absorption}\\
$\mathrm{60\%CO_2-40\%N_2}$ & 2.57$\times10^4$ &$\sim$2.6$\times10^4$ &293.5 &\cite{ejakov2003acoustic}\\
$\mathrm{20\%CO_2-80\%O_2}$ & 1.26$\times10^4$ &$\sim$1.2$\times10^4$ &300    &\cite{bass1973vibrational}\\
$\mathrm{30\%CO_2-70\%O_2}$ & 1.55$\times10^4$ &$\sim$1.8$\times10^4$ &300    &\cite{bass1973vibrational}\\
$\mathrm{98\%CO_2-2\%air}$ & 4.03$\times10^4$ &$\sim$4.0$\times10^4$ &298    &\cite{petculescu2006prototype}\\
$\mathrm{98\%CH_4-2\%air}$ & 1.12$\times10^5$  &$\sim$1.3$\times10^5$ &298    &\cite{petculescu2006prototype}\\
\hline
\end{tabular}
\end{table}

For many pure gases and gas mixtures, only one dominant relaxation process occurs in acoustic relaxation, and the absorption spectrum show only one spectral peak with the maximum absorption value along the abscissa (log $f/P$). The position of the spectral peak is determined by the relaxation frequency which is the reciprocal of molecular relaxation time.\cite{petculescu2012quantitative,hu2014acoustic} As a result, the calculation of the relaxation frequencies is the critical step in quantitative description of sound relaxation. Here we compare the relaxation frequencies from our proposed model with the experimental data for various gas compositions as listed in Table 2. Obviously, the relaxation frequencies from theoretical calculation are consistent with experimental results. The comparisons of relaxation frequencies further verify our proposed model.

\subsection{Comparisons with parallel relaxation theory}

Among the previous relaxation methods, Zhang's model\cite{ke2013decoupling} is one of the parallel relaxation method for gas mixtures. In figure 4, the comparison between our proposed model and Zhang's model is displayed for the gas mixtures $\mathrm{CO_2-N_2}$ with different molar fractions. The theoretical curves from our model (solid curve) and Zhang's model (dashed curve) both match with the experimental results from Ejakov.\cite{ejakov2003acoustic} The spectra from our work show slightly larger amplitudes than the curves of Zhang's model and more consistent with the experimental data. The amplitude deviations are primarily attributed to the different relaxation energy transfer mechanisms between our model and Zhang's model. In our model, the vibrational energy of a certain gas component is considered to relax via the lowest vibrational mode. While in Zhang's model, the vibrational energy is assigned to each mode of gas mixture and the separation of relaxing energy results in the lower spectral amplitude. In these cases (only one dominant relaxation process), the deviations between spectral peaks from our proposed model and Zhang's model are approximately $5\%$ and in an acceptable range. 

\begin{figure} [htbp]
\label{figure4}
\centering
\includegraphics
[width=0.8\textwidth,height=0.65\textwidth]{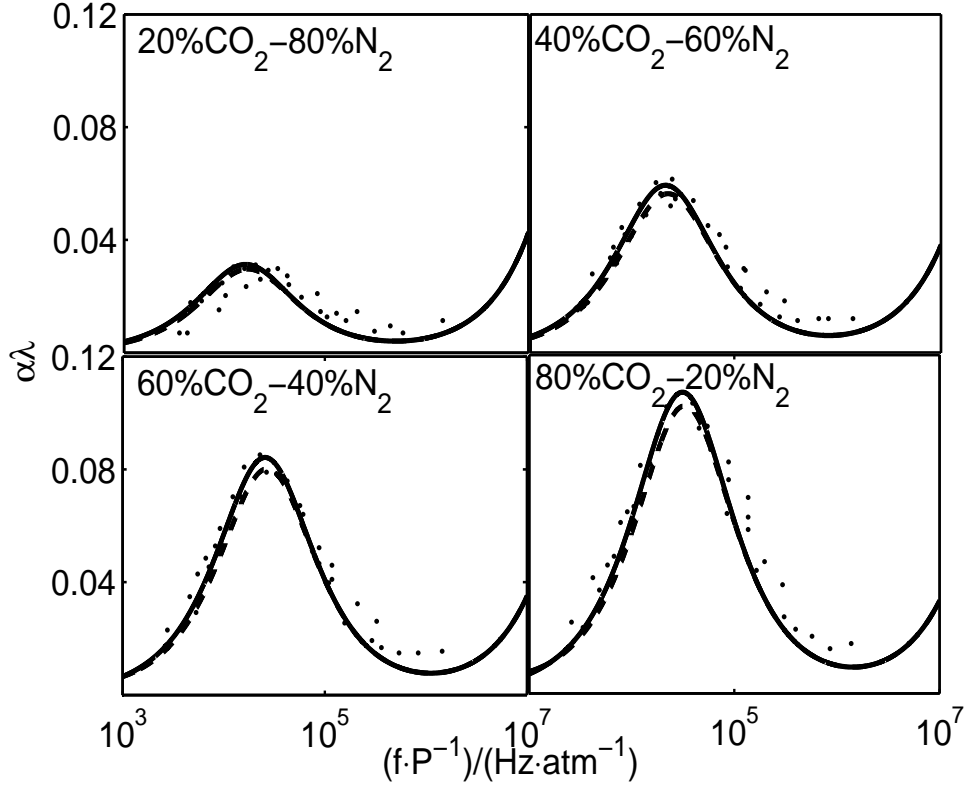}
\caption{\label{fig:epsart} Comparisons of theoretical acoustic absorption (relaxation and classical absorption) spectra predicted from our work (solid curves), theoretical spectra from Zhang's model\cite{ke2013decoupling} (dashed curves) and experimental data from Ejakov\cite{ejakov2003acoustic} (dots) for $\mathrm{CO_2-N_2}$ mixtures with 20$\%$, 40$\%$, 60$\%$, 80$\%$ $\mathrm{CO_2}$ at temperatures of 292.6, 293.7, 293.5 and 294.0 K, respectively.}
\end{figure}

However, our proposed model is more applicable than Zhang's model for gas mixtures with multiple dominant relaxation processes. In figure 5, the theoretical spectra from our proposed model and Zhang's model are presented for the gas mixture of $\mathrm{CO_2-O_2}$ at 450 K as well as experimental data from Bass.\cite{bass1973vibrational} In this case, the gas mixture of $\mathrm{CO_2-O_2}$ shows two clear spectral peaks. And the curve from our model is consistent with experimental results while the curve from Zhang's model shows one spectral peak. The comparison demonstrates that our proposed model obtains more reliable acoustic relaxation absorption spectra. Therefore, the proposed model could be used to predict acoustic relaxation absorption with higher accuracy.

\begin{figure} [htbp]
\label{figure5}
\centering
\includegraphics
[width=0.8\textwidth,height=0.65\textwidth]{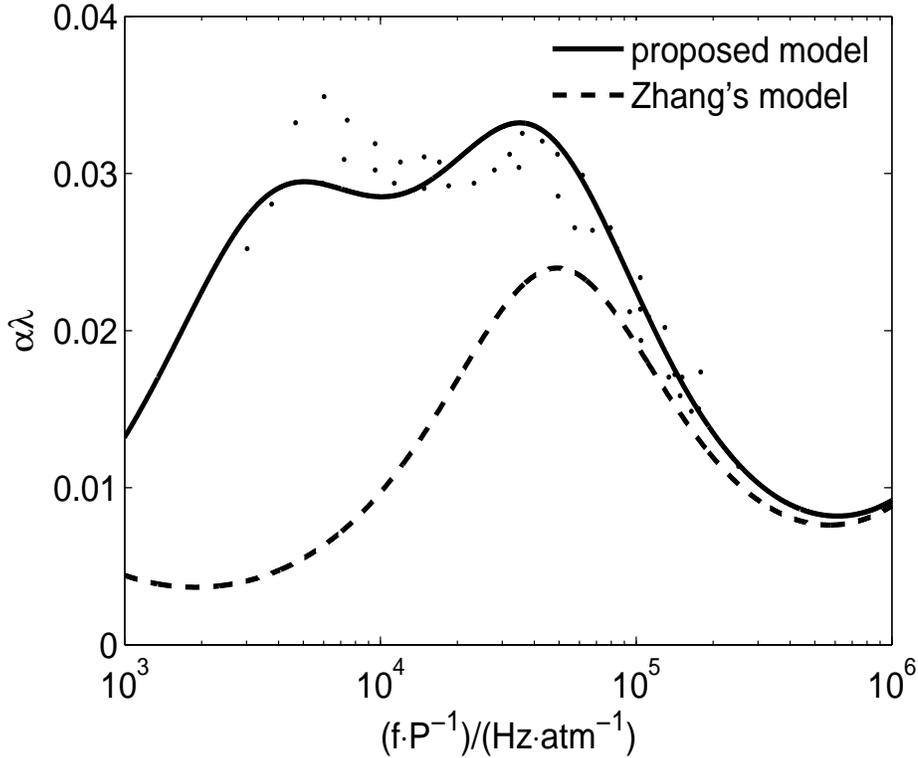}
\caption{\label{fig:epsart} Comparison of theoretical acoustic absorption (relaxation and classical absorption) spectra predicted from our proposed method (solid curve), spectra from Zhang's model (dashed curve), and experimental data from Bass\cite{bass1973vibrational} (dots) for $\mathrm{10\%CO_2-90\%O_2}$ at 450 K. }
\end{figure}

In addition, the effects of energy transfer between vibrational modes in gas mixtures $\mathrm{CO_2-O_2}$ of figure 5 and $\mathrm{CO_2-N_2}$ of figure 4 are quite different. Based on the proposed energy transfer mechanism, the rate-determining V-V process between the lowest mode of $\mathrm{O_2}$ 1554 cm$^{-1}$ and the vibrational mode of $\mathrm{CO_2}$ 1388 cm$^{-1}$ plays significant role in vibrational energy transfer of the molecular relaxation of $\mathrm{CO_2-O_2}$ and greatly influences the effective relaxation time calculated from equations (2.12)-(2.15). In contrast, none of the V-V processes is rate-determining and taken into consideration for the molecular relaxation of $\mathrm{CO_2-N_2}$ in the proposed model. Despite the differences in energy transfer processes of molecular relaxation and the gas temperature conditions (T=450K for $\mathrm{CO_2-O_2}$ of figure 5 and T=293K for $\mathrm{CO_2-N_2}$ of figure 4), the good agreements of the theoretical spectra in figure 4 and 5 with the experimental results verify the reliability of the proposed model.

\subsection{Relaxation contribution of gas composition}

From the preceding gas examples in figures 2 to 5, our proposed model obtains the acoustic absorption spectra in good agreement with experimental data. Furthermore, our proposed model reveals a new insight for participation mechanism of gas composition in molecular acoustic relaxation, which is investigated in this section.

In the proposed model, molecular relaxation of gas mixtures is achieved by individual energy transfer processes of the containing gas compositions. The individual energy transfer process of each gas composition appears as single relaxation process in the molecular relaxation of mixture, which can be considered as the relaxation contribution of each gas composition. From effective specific heats of gas mixtures in equation (2.11), we extract the external specific heat of mixture, internal specific heat and effective relaxation time of composition $k$ to describe the relaxation contribution of composition $k$. Hence, the effective specific heat of gas mixtures containing M components can be decomposed into M parts of the following form as,
\begin{equation} \label{eq18}
C_k^{eff}=C_m^\infty+\frac{a_k C_k^{int}}{1+i\omega \tau_k^{eff}},
\end{equation}
In equation (3.1), the single relaxation process with relaxation time $ \tau_k^{eff}$ represents the deexcitation path provided by corresponding gas composition $k$. The decomposed effective specific heat in equation (3.1) is considered as the specific heat contribution of the corresponding gas composition. Substituting the decomposed effective specific heat in equation (3.1) into equations (2.16) and (2.17), the corresponding decomposed relaxation absorption spectrum $\mu=\alpha_r\lambda$  can be obtained.\cite{zhu2016capturing}

Here we take an example of gas mixture $\mathrm{10\%CO_2-90\%O_2}$  at 600 K to illustrate the relaxation contributions of gas compositions. The molecular relaxation of $\mathrm{CO_2-O_2}$ is decomposed into two single relaxation processes corresponding to relaxation contributions of $\mathrm{CO_2}$ and $\mathrm{O_2}$, respectively. In figure 6, the decompositions of effective specific heat curves and relaxation absorption spectra are demonstrated. Compared with the decomposed effective specific heats showing semicircle curves, the decomposed absorption spectra provide more intuitive information. The mixture's spectrum shows two obvious peaks and it is decomposed into the contribution spectra of $\mathrm{CO_2}$ and $\mathrm{O_2}$, respectively. As figure 6(b) shows, $\mathrm{O_2}$ is the main contributor to the spectral peak at low frequency while $\mathrm{CO_2}$ mainly contributes to the peak at high frequency. Therefore, the contributions of gas compositions can be extracted from the decomposition based on our proposed model and be clearly demonstrated by the decomposed spectra.

\begin{figure} [htbp]
\label{figure6}
\subfloat[effective specific heat]{
\includegraphics
[width=0.5\textwidth,height=0.4\textwidth]{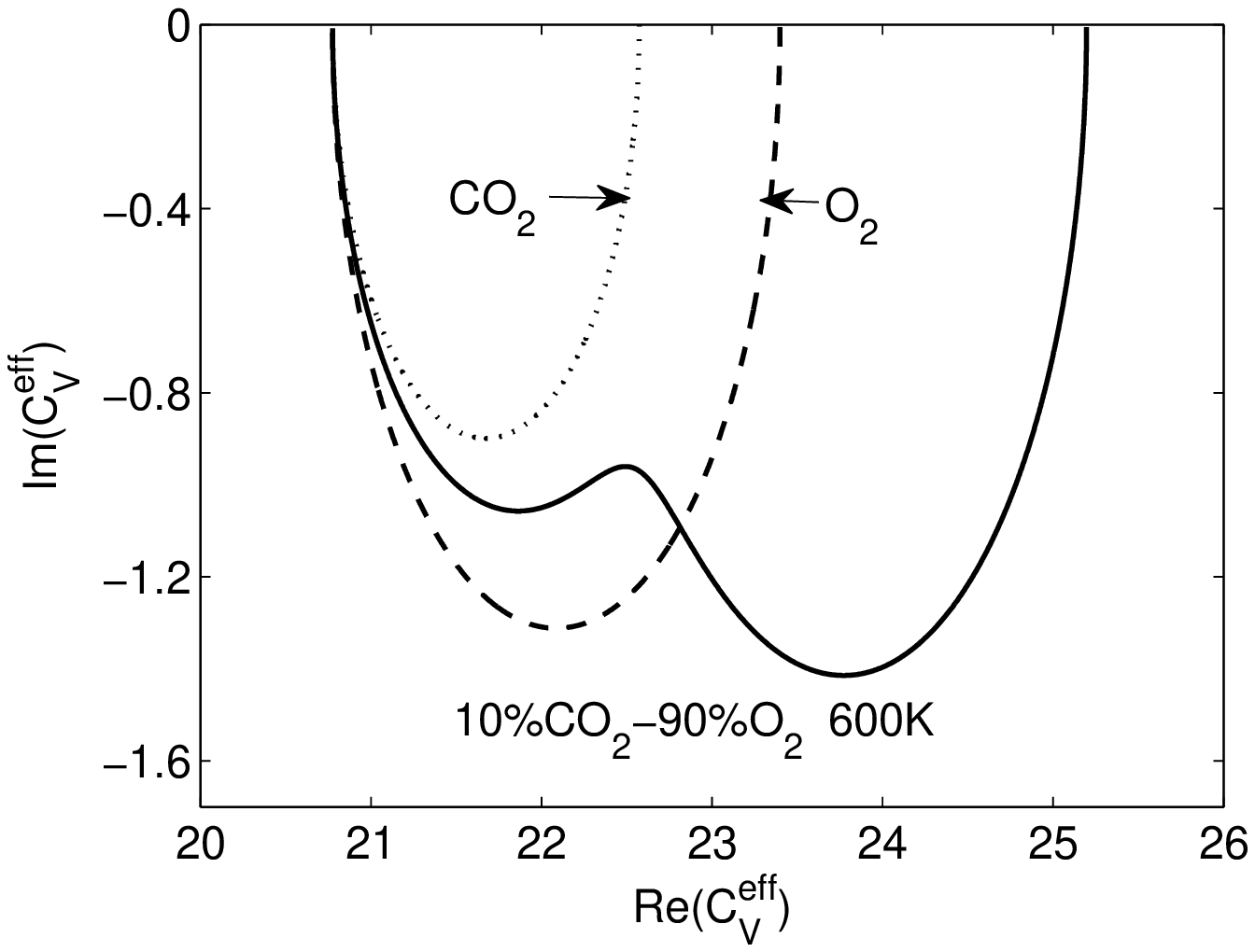}
}
\subfloat[acoustic relaxation absorption spectrum]{
\includegraphics
[width=0.5\textwidth,height=0.4\textwidth]{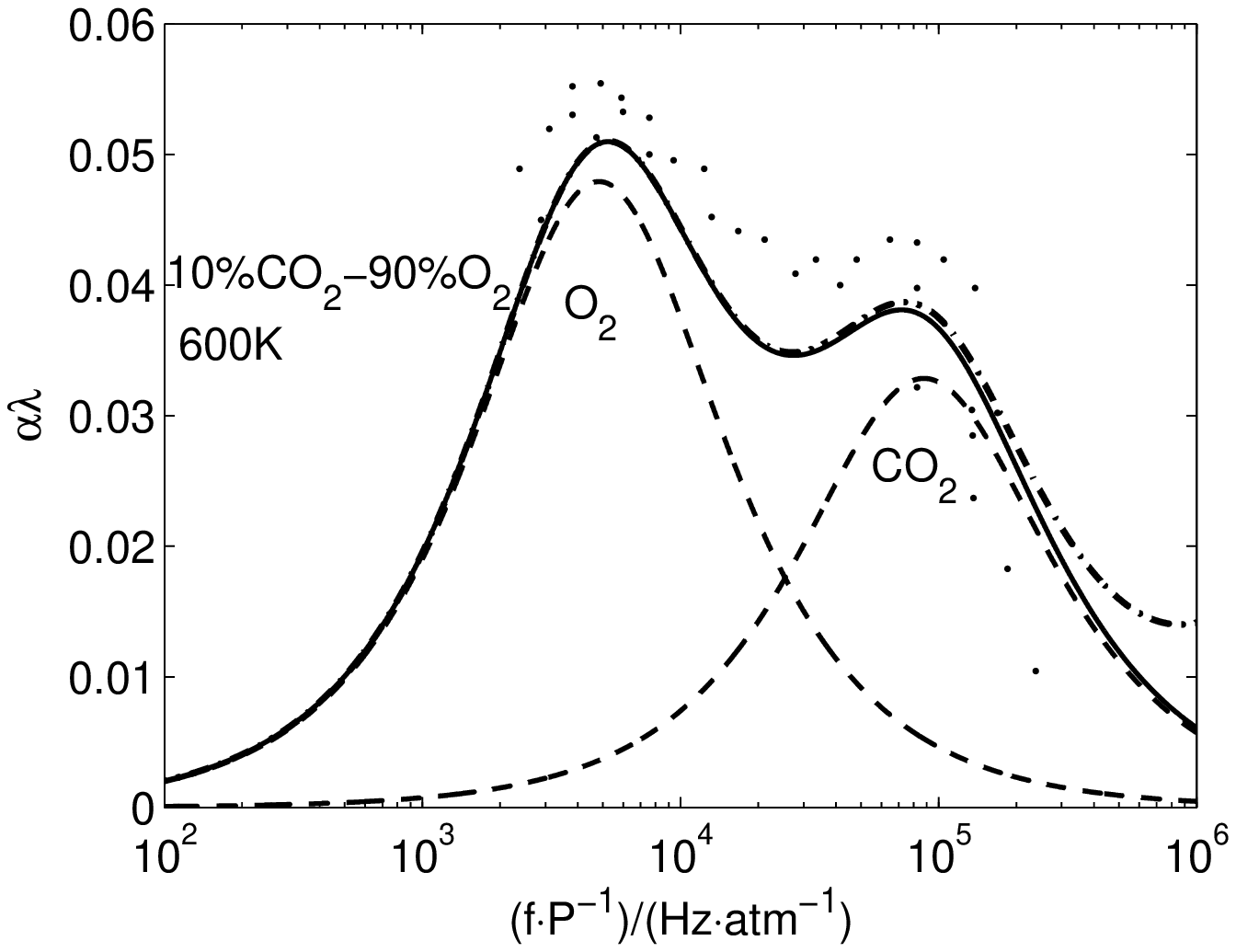}
}
\caption{Decomposition of (a) effective specific heat and (b) acoustic relaxation absorption spectrum for participation contributions of gas compositions in mixture $10\%\mathrm{CO_2}-90\%\mathrm{O_2}$ at 600 K. Solid curves: Theoretical effective specific heat traces in (a) and relaxation absorption spectra in (b) predicted by our model for $\mathrm{CO_2}-\mathrm{O_2}$. Dotted curves: Decomposed results for $\mathrm{CO_2}$. Dashed curves: Decomposed results for $\mathrm{O_2}$.  Dots: Experimental data from Bass\cite{bass1973vibrational}. Dash-dotted curve in (b): theoretical acoustic absorption (relaxation and classical absorption) spectra.}
\end{figure}

Compared with traditional parallel relaxation theory, our proposed model reveals the contributions of gas compositions to acoustic relaxation in gases. The parallel relaxation theory takes the energy transfer between all the vibrational modes into account and couples vibrational energy of all different vibrational modes. Each single relaxation process of parallel relaxation theory contains the coupling contributions of all the molecular vibrational modes.\cite{ke2013decoupling,zhang2017calculating} In our proposed model, each gas composition contributes a single relaxation process to molecular relaxation and the participation contribution of gas composition can be demonstrated by the decomposed spectra of the single relaxation processes. Hence, the participation mechanism of gas compositions in molecular relaxation becomes much clearer.

\section{Conclusion}

In this work, we have developed the valid analytical model for acoustic relaxation absorption in gases to clarify the relaxation contributions of gas compositions. By combining the parallel and series relaxation theory, the energy transfer of molecular relaxation is characterized with V-T and rate-determining V-V processes of the lowest vibrational modes. The model reveals that molecular relaxation of pure gases is realized by one single relaxation process due to the major role of the lowest mode in energy transfer, and molecular relaxation of gas mixtures is achieved by the individual energy transfer process of each component. The acoustic absorption spectra obtained from our proposed model are in good agreement with experimental data, which are more reliable for gas mixtures with multiple dominant relaxation processes compared with previous models. Furthermore, the model reveals that each gas composition provides a deexcitation path in molecular relaxation and demonstrates the relaxation contributions of the gas compositions by the decomposed spectra of single relaxation processes. Our model holds great potential for practical detection of gas compositions through extracting gas composition contributions from mixtures.

\section*{Acknowledgement}

This work was supported by the National Natural Science Foundation of China (Grant No. 61371139, 61571201 and 61461008).








\end{document}